\documentclass{llncs}
\usepackage{stmaryrd}
\usepackage{amsmath}
\usepackage{epstopdf}
\usepackage{amssymb}
\usepackage{pifont}
\usepackage[nocompress]{cite}
\usepackage{caption}
\usepackage[vlined,ruled,linesnumbered]{algorithm2e}
\usepackage{graphicx}
\usepackage{array}
\usepackage{listings}
\usepackage[table]{xcolor}

\begin{document}

\title{DroidGen: Constraint-based and Data-Driven Policy Generation for Android}

\author{Mohamed Nassim Seghir\inst{1} \and  David Aspinall\inst{2}}
\institute{University College London \and University of Edinburgh}

\maketitle
\begin{abstract}
We present DroidGen a tool for automatic anti-malware policy
inference. DroidGen is data-driven: uses a training set of malware and benign applications and makes call to a constraint solver to generate a
policy under which a maximum of malware is excluded and a maximum
of benign applications is allowed. Preliminary results are encouraging.
We are able to automatically generate a policy which filters out 91\%
of the tested Android malware. Moreover, compared to black-box machine
learning classifiers, our method has the advantage of generating policies
in a declarative readable format. We illustrate our approach, describe its
implementation and report on experimental results.

\end{abstract}

\section{Introduction}

Security on Android is enforced via permissions giving access to 
resources on the device. These permissions are often too coarse and 
their attribution is based on an all-or-nothing decision in the vast 
majority of Android versions in actual use. Additional security policies can be 
prescribed to impose a finer-grained control over resources. However, 
some key questions must be addressed: who writes the policies? What is 
the rational behind them? An answer could be that policies are written 
by experts based on intuition and prior knowledge. What can we do then 
in the absence of expertise? Moreover, are we sure that they provide enough coverage?

We present DroidGen a tool for the systematic generation of anti-malware policies. DroidGen is fully automatic and data-driven: it takes as input two training sets of benign and malware applications and returns a policy as output. The resulting policy represents an optimal solution for a constraint satisfaction problem expressing that the discarded malware should be maximized while the number of excluded benign applications must be minimized. The intuition behind this is that the solution will capture the maximum of features which are specific to malware and less common to benign applications. Our goal is to make the generated policy as general as possible to the point of allowing us to make decisions regarding new applications which are not part of the training set.  

In addition to being fully push-button, DroidGen is able to generate a policy that filters out 91\% of malware from a representative testing set of Android applications with only a false positive rate of 6\%. Moreover, having the policies in a declarative readable format can boost the effort of the malware analyst by providing diagnosis and pointing her to suspicious parts of the application.
In what follows we present the main ingredients of DroidGen, describe their functionality and report on experimental results.

\section{Application Abstraction}
\begin{figure}
  \begin{center}
    \includegraphics[scale=.4]{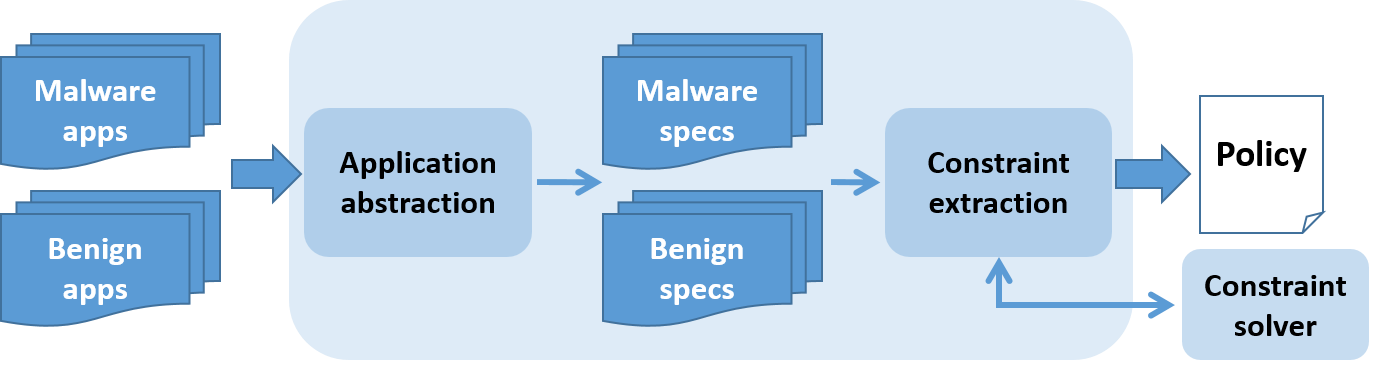}
    \caption{Illustration of DroidGen's Main Ingredients}
    \label{fig:droidgen}
    \vspace{-.5cm} 
  \end{center}
\end{figure}

DroidGen proceeds in several phases: application abstraction, constraint extraction and constraint solving, see Figure~\ref{fig:droidgen}.
Our goal is to infer policies that distinguish between good and bad behaviour. As it is not practical to have one policy per malicious application, we need to identify common behaviours of applications. Hence the first phase of our approach is the derivation of specifications (abstractions) which are general representations of applications. Given an application $A$, the corresponding high level specification $\mathsf{Spec}(A)$ consists of a set of properties $\{p_1,\ldots,p_k\}$ such that each property $p$ has the following grammar:
\[
\begin{array}{rcl}
 p&:= &  c:r\\
c&:=&\bf{\textsf{entry\_point $|$ activity $|$ service $|$ receiver}}\\
& & \textsf{$|$ onclick\_handler $|$ ontouch\_handler $|$\;} lc\\
lc&:=&\bf{\textsf{oncreate $|$ onstart $|$ onresume $|$ \ldots}}\\
u &:=& perm \;|\; api\\
\end{array}
\]
A property $p$ describes a context part $c$ in which a resource $r$ is used. The resource part can be either a permission \emph{perm} or \emph{api} referring to an api method identifier which consists of the method name, its signature and the class it belongs to. The context $c$ can be \textsf{entry\_point} referring to all entry points of the app, \textsf{activity} representing methods belonging to activities, \textsf{service} for methods belonging to service components\footnote{\textsf{activity}, \textsf{service} and \textsf{receiver} are some of the building blocks of Android applications.}, etc. We also have \textsf{onclick\_handler} and \textsf{ontouch\_handler} respectively referring to click and touch event handlers. Moreover, $c$ can be an activity life-cycle callback such as \textsf{oncreate}, \textsf{onstart}, etc.\footnote{Some components have a life-cycle governing their callbacks invocation.} Activity callbacks as well as the touch and click event handlers are also entry points. 

A property $p$ of the form $c:r$ belongs to the specification of an application $A$ if $r$ ($perm$ or $api$) is used within the context $c$ in $A$. In other words: it exists at least one method matching $c$ from which $r$ is transitively called (reachable). To address such a query, we compute the transitive closure of the call graph \cite{SeghirA15}. We propagate permissions (APIs) backwards from callees to callers until we reach a fixpoint.

For illustration, let us consider the example in Figure~\ref{fig:recorder}. On the left hand side, we have code snippets representing a simple audio recording application named \textsf{Recorder} which inherits from an Activity component. On the right hand side, we have the corresponding specifications in terms of APIs (Figure~\ref{fig:recorder}(a)) and in terms of permissions (Figure~\ref{fig:recorder}(b)). The method \textsf{setAudioSource}, which sets the recording medium for the media recorder, is reachable (called) from the Activity life-cycle method \textsf{onCretae}, hence we have the entry \textsf{oncreate: setAudioSource} in the specification map (a). We also have the entry \textsf{oncreate}: \textsc{record\_audio} in the permission-based specification map (b) as the permission \textsc{record\_audio} is associated with the API method \textsf{setAudioSource} according to the Android framework implementation. Similarly, the API method \textsf{setOutputFile} is associated with the context \textsf{onclick} (a) as it is transitively reachable (through \textsf{startRecording}) from the click handler method \textsf{onClick}. Hence permission \textsc{write\_external\_storage}, for writing the recording file on disk, is also associated with \textsf{onclick} (b). Both APIs and permissions are also associated with the context \textsf{activity} as they are reachable from methods which are activity members. We use results from \cite{AuZHL12} to associate APIs with the corresponding permissions. 
\begin{figure}
\begin{center}
\begin{tabular}{c@{\hspace{0.2in}}c}
\begin{lstlisting}
public class Recorder extends Activity 
  implements OnClickListener{ 	 
  private MediaRecorder myRecorder;  
  ...
  public void onCreate(...) {
	myRecorder = new MediaRecorder();	
	// uses RECORD_AUDIO permission
	myRecorder.setAudioSource(...);
	}
        
  private void startRecording() {        
	// uses WRITE_EXTERNAL_STORAGE	
	myRecorder.setOutputFile(...);	
	recorder.start();        
  	}
  
  public void onClick(...) {        
	startRecording();        		
  	}
}
\end{lstlisting}
&
\scriptsize{
	\begin{tabular}{c}
			\textsf{Spec}(Recorder)
			\\	
			\hline\hline
			\begin{tabular}{rl}
			  \textsf{oncreate}:&	setAudioSource \\	
		    	  \textsf{onclick}:&	setOutputFile \\
		    	  \textsf{activity}:&	setAudioSource \\	
		    	  \textsf{activity}:&	setOutputFile \\		    			    		    
			\end{tabular}
			\\
			\hline	
			(a)
			\\
			\\
			\textsf{Spec}(Recorder)
			\\	
			\hline\hline
			\begin{tabular}{rl}
			  \textsf{oncreate}:&	 \textsc{record\_audio}\\	
		    	  \textsf{onclick}:&	\textsc{write\_external\_storage} \\
		    	  \textsf{activity}:&	\textsc{record\_audio} \\	
		    	  \textsf{activity}:&	\textsc{write\_external\_storage} \\
			\end{tabular}
			\\
			\hline	
			(b)						
	\end{tabular}
	}
\end{tabular}

\end{center}
\caption{Code snippets sketching a simple audio recording application together with the corresponding specifications based on APIs (a) and based on permissions (b)}
\label{fig:recorder}
\end{figure}

\section{Specifications to Policies: an Optimisation Problem?}
DroidGen tries to derive a set of rules (policy) under which a maximum number of benign applications is allowed and a maximum of malware is excluded. This is an optimization problem with two conflicting objectives. Consider
\begin{figure}[h!]
\vspace{-.5cm}  
\begin{center}
\begin{tabular}{c@{\hspace{0.5in}}|@{\hspace{0.5in}}c}
\begin{tabular}{c@{\hspace{0.1in}}c@{\hspace{0.1in}}c}
$\mathsf{Spec}(benign_1)$ &=& $\{p_{a}\}$\\
\hline
$\mathsf{Spec}(benign_2)$ &=& $\{p_{c}\}$\\
\hline
$\mathsf{Spec}(benign_3)$ &=& $\{p_{b}, p_{e}\}$\\
\end{tabular}
&
\begin{tabular}{c@{\hspace{0.1in}}c@{\hspace{0.1in}}c}
$\mathsf{Spec}(malware_1)$ &=& $\{p_{a}, p_{b}\}$\\
\hline
$\mathsf{Spec}(malware_2)$ &=& $\{p_{a}, p_{c}\}$\\
\hline
$\mathsf{Spec}(malware_3)$ &=& $\{p_{d}\}$\\
\end{tabular}
\end{tabular}
\end{center}
\label{fig:specs}
\vspace{-.7cm}
\end{figure}

\noindent Each application (benign or malware) is described by its specification consisting of a set of properties ($p_i$'s). As seen previously, a property $p_i$ can be for example \textsf{activity : record\_audio}, meaning that the permission \textsf{record\_audio} is used within an activity. A policy excludes an application if it contradicts one of its properties. We want to find the policy that allows the maximum of benign applications and excludes the maximum of malware. This is formulated as: 
\[
Max [\underbrace{I(p_a) + I(p_c) + I(p_b \wedge p_e)}_{benign} - \underbrace{(I(p_a \wedge p_b) + I(p_a \wedge p_c) + I(p_d))}_{malware}]
\]
where $I(x)$ is the function that returns 1 if $x$ is true or 0 otherwise. This type of optimization problems where we have a mixture of theories of arithmetic and logic can be efficiently solved via an SMT solver such as Z3 \cite{MouraB08}. It gives us the solution: $p_a = 0$, $p_b = 1$, $p_c = 1$, $p_d = 0$ and $p_e = 1$. Hence, the policy will contain the two rules $\neg p_a$ and $\neg p_d$ which filter out all malware but also exclude the benign application $benign_1$. A policy is violated if one of its rules is violated.
\paragraph{\bf Policy Verification and Diagnosis.}
Once we have inferred a policy, we want to use it to filter out applications violating it. A policy $P = \{\neg p_1,\ldots,\neg p_k\}$ is violated by an application $A$ if $\{p_1,\ldots,p_k\} \cap \mathsf{Spec}(A) \neq \emptyset$, meaning that $A$ contradicts (violates) at least one of the rules of $P$. In case of policy violation, the violated rule, e.g. $\neg p$, can give some indication about a potential malicious behaviour. DroidGen maps back the violated rule to the code in order to have a view of the violation origin. For $p = (c:u)$, a sequence of method invocations $m_1,..,m_k$ is generated, such that $m_1$ matches the context $c$ and $m_k$ invokes $u$.  
\section{Implementation and Experiments}
\label{sec:experiments}
DroidGen\footnote{www0.cs.ucl.ac.uk/staff/n.seghir/tools/DroidGen} is written in Python and uses Androguard\footnote{https://github.com/androguard} as front-end for parsing and decompiling Android applications. DroidGen automatically builds abstractions for the applications which are directly accepted in APK binary format. This process takes around 6 seconds per application. An optimization problem in terms of constraints over the computed abstractions is then generated and the Z3 SMT solver is called to solve it. Finally, the output of Z3 is interpreted and translated to a readable format (policy). Policy generation takes about 7 seconds and its verification takes no more than 6 seconds per app on average. 

We derived two kinds of policies based on a training set of 1000 malware applications from Drebin\footnote{\textsf {http://user.informatik.uni-goettingen.de/$\sim$darp/drebin/}} and 1000 benign ones obtained from Intel Security (McAfee). The first policy $P_p$ is solely based on permissions and is composed of 65 rules. The other policy $P_a$ is exclusively based on APIs and contains 152 rules. Snippets from both policies are illustrated in the appendix. We have applied the two policies to a testing set of 1000 malware applications and 1000 benign ones (different from the training sets) from the same providers. 
Results are summarised in Table~\ref{tab:results}.
\vspace{-.3cm}
\begin{table}[!h]
\begin{center}
\begin{tabular}{|c|c|c|}
\hline
Policy & Malware filtered out & Benign excluded\\
\hline
\hline
APIs ($P_a$)& 910/1000 & 59/1000\\ 
Permission ($P_p$) & 758/1000 & 179/1000\\
\hline
\end{tabular}  
\caption{Results for a permissions-based policy ($P_p$) vs. an API-based one ($P_a$)} 
\label{tab:results}
\vspace{-1cm}
\end{center}
\end{table}
The policy $P_a$ composed of rules over APIs performs better than the one that uses permissions in terms of malware detection as it is able to filter out $91\%$ of malware while $P_p$ is only able to detect $76\%$. It also has a better false positive rate as it only excludes $6\%$  of benign applications, while $P_p$ excludes $18\%$. 
Being able to detect $91\%$ of malware is encouraging as it is comparable to the results obtained with some of the professional security tools (https://www.av-test.org/)\footnote{We refer to AV-TEST benchmarks dated September 2014 as our dataset was collected during the same period.}. Moreover, our approach is fully automatic and the actual implementation does not exploit the full expressiveness of the policy space as we only generate policies in a simple conjunctive form. We plan to further investigate the generation of policies in arbitrary propositional forms.

\section{Related Work} 
\label{sec:related}
Many tools for analysing various security aspects of Android have emerged \cite{flowdroid, FahlHMSBF12, ChinFGW11, BackesGHMS13}. They either check or enforce certain security properties (policies). These policies are either hard-coded or manually provided. Our work complements such tools by providing the automatic means for inferring the properties to be checked. Hence, DroidGen can serve as a front-end for a verification tool such as EviCheck \cite{SeghirA15} to keep the user completely out of the loop.

Also various machine-learning-based approaches have been proposed for malware detection \cite{YangXGYP14, ArpSHGR14, AvdiienkoKGZARB15, YangXALXE15}. While some of them outperform our method, We did not yet exploit the entire power of the policy language. We are planning to allow more general forms for the rules, which could significantly improve the results. Moreover, many of the machine-learning based approaches do not provide further indications about potential malicious behaviours in an application. Our approach returns policies in a declarative readable format which is helpful in terms of code inspection and diagnosis. Some qualitative results are reported in the appendix. To the best of our knowledge, our approach is unique for being data-driven and using a constraint solver for inferring anti-malware policies.

\section{Conclusion and Future Work} 
\label{sec:learn_pol}
We have presented DroidGen a tool for the automatic generation of anti-malware policies. It is data-driven and uses a constraint solver for policy inference. DroidGen is able to automatically infer an anti-malware policy which filters out 91\% of the tested malware with the additional benefit of being fully automatic. Having the policies in declarative readable format can boost the effort of the malware analyst by pointing her to suspicious parts of the application. As future work, we plan to generate more expressive policies by not restring their form to conjunctions of rules. We also plan to generate anti-malware policies for malware families with the goal of obtaining semantics-based signatures (see appendix).

\bibliographystyle{abbrv}
\bibliography{biblio}
\end{document}